\documentclass{article}

\usepackage{PRIMEarxiv}

\usepackage[utf8]{inputenc}
\usepackage[T1]{fontenc}
\usepackage[numbers,sort&compress]{natbib}
\usepackage{hyperref}
\usepackage{url}
\usepackage{xurl}
\usepackage{booktabs}
\usepackage{amsfonts}
\usepackage{amsmath}
\usepackage{nicefrac}
\usepackage{microtype}
\usepackage{fancyhdr}
\usepackage{graphicx}
\usepackage{multirow}
\usepackage{array}
\usepackage[most]{tcolorbox}
\usepackage{tabularx}
\usepackage{tikz}
\usetikzlibrary{arrows.meta,positioning,fit}

\usepackage{titlesec}
\titlespacing*{\section}{0pt}{1.2ex}{0.8ex}

\title{Contextualization of Third-Party \\ Cloud Security Findings}

\author{
  \textbf{Leon Goldberg},
  \textbf{Gal Engelberg}\\
  Sola Security, Tel Aviv, Israel \\
  \texttt{Corresponding author: gal.e@sola.security}
}

\hypersetup{
pdftitle={Contextualization of Third-Party Cloud Security Findings},
pdfauthor={Leon Goldberg, Gal Engelberg},
pdfkeywords={cloud security, contextualization, severity, asset graph, deep research agent},
}

\begin{document}
\maketitle

\begin{abstract}
Finding severity is the main driver of how security teams prioritize remediation. For third-party cloud security findings, that severity is static: the rule that raised the finding assigns it before the rule meets any environment, so it reflects the risk of the condition in general rather than the risk the finding poses to the concrete environment where it lives. Scoring standards define where environment-specific context belongs. How far that context changes finding severities in production, where the deciding evidence lies, and whether it holds against the live environment have not been measured. We address this gap with \emph{contextualization}, re-deriving each finding's severity from evidence in the environment where the finding lives. A deep research agent over a precomputed cross-signal asset graph investigates each finding against the resource's state, its graph neighborhood, and other products' signals, and returns an adjusted severity with an evidence trace. We evaluate it in a production field study of 9,967 vendor HIGH findings from two commercial cloud security platforms across eight real production environments, on three criteria: the faithfulness of the facts behind each verdict to the live environment, the dependence of each decision on context beyond the flagged resource, and the regularity of the reasoning. Three in four findings are re-graded, mostly downward, and the same rule often moves in opposite directions inside a single environment. About half of the decisive evidence lies beyond the flagged resource, and read-only probes of live infrastructure confirm the decisive fact for 99.4\% of decided findings.
\end{abstract}


\section{Introduction}

Cloud security posture and workload protection tools, products that evaluate cloud resources against rules and report each violation as a finding, are the main source of the work queues that cloud security teams triage. Severity is what orders those queues. Vulnerability management programs rank remediation mostly on severity scores and expert judgment~\citep{jacobs2021epss}, and security operations center (SOC) incident queues are often sorted by coarse severity~\citep{freitas2026adaptive}. Yet the severity of a third-party finding is static. The rule that raised the finding assigns it before the rule meets any environment: across all production environments we examined, each of 599 vendor rules emits exactly one severity over 99,073 failing findings of every grade, whatever resource or environment it fires on. A static severity reflects the risk of the condition in general. It does not reflect the risk the finding poses to the concrete environment where it lives, and a queue ordered by it is ordered by the general case. The scoring standards reserve a place for the missing context: the Common Vulnerability Scoring System (CVSS) defines environmental metrics that adjust a base score to a particular deployment~\citep{first2019cvss31,first2023cvss40}, the Stakeholder-Specific Vulnerability Categorization (SSVC) assesses the exposure of the specific system as deployed~\citep{spring2021ssvc}, and the National Institute of Standards and Technology (NIST) risk assessment conditions likelihood on predisposing conditions and on the controls in place~\citep{nist2012sp80030}. A static severity leaves that place empty. Analysts report the consequence: in a qualitative study of SOC practitioners, most alarms that practitioners perceived as false positives were correct detections of activity that the organization's own environment made legitimate~\citep{alahmadi2022false}.

A rule carries the context its author can encode once for every environment: the condition, the resource type, and the typical consequence. The context that decides what the condition is worth in a given environment lies elsewhere: whether the resource is attached to anything, whether a workload is running, whether a network path from the internet exists, whether an account-level control already blocks the exposure, and what another product reports about the same resource. We call the difference between the severity a rule assigns and the severity the evidence in a given environment supports \emph{the rule-severity gap}. Commercial platforms build environment graphs and attack-path views~\citep{microsoft2026attackpath,liu2025wiztoxic}, and controlled studies report that rule-based posture alerts include many false positives that context or active validation removes~\citep{dikshant2025precision,dikshant2026alertnoise}. Structured context also raises the accuracy of language models on security tasks, for identity reasoning~\citep{yavin2026ispm} and Kubernetes remediation~\citep{shaikh2026runtime}, and a security context layer reaches higher coverage than coding agents on cloud investigation tasks~\citep{goldberg2026coding}. The size of the rule-severity gap on production findings, the location of the evidence that closes it, and whether that evidence holds against live infrastructure have not been measured.

We present a production field study of the \emph{contextualization} of third-party findings: re-deriving each finding's severity from evidence in the environment where the finding lives, and recording that evidence. The system under study has two components. An \emph{asset graph} reconciles the signals of every connected security product into one graph of resources, controls, and identities, and precomputes the structural facts that decide risk, such as reachability and attachment~\citep{engelberg2026ai}. A \emph{contextualization agent} is a deep research agent over that graph and over the tabular data of every connected product, built on the investigation layer of~\citet{goldberg2026coding}: for each finding it plans an investigation, issues structured and graph queries, searches for tables held by other products, and ends with a verdict that may move the severity, together with a trace of the evidence it used. We apply the system to the HIGH queue of two commercial cloud security platforms, Platform A and Platform B, across eight real production environments. Both platforms appear among the recognized leaders in Gartner's and Forrester's evaluations of cloud-native application protection platforms (CNAPP) and cloud workload protection platforms (CWPP).

The contribution of this paper is a conceptual architecture for an agent that contextualizes findings. It specifies the agent's unit of investigation, the evidence actions available to it, its reasoning loop, and an evidence policy that sets a higher bar for raising a severity than for lowering it. We evaluate the architecture in production on 9,967 vendor HIGH findings against three criteria: \emph{faithfulness}, whether the facts its verdicts rest on hold in the live environment; \emph{context dependence}, how much of each decision needs evidence beyond the flagged resource; and \emph{regularity of reasoning}, whether the verdicts follow a small set of recurring patterns that independent LLM judges identify reproducibly.

The rest of the paper is structured as follows. Section~\ref{sec:related} positions the work against scoring standards, reachability analysis, and language-model and industrial agents; Section~\ref{sec:agent} describes the contextualization agent; Section~\ref{sec:evaluation} describes the corpus and the evaluation; Section~\ref{sec:results} reports the results; Sections~\ref{sec:discussion} and~\ref{sec:conclusion} discuss implications and limitations and conclude.

\section{Related Work}
\label{sec:related}

Prior work uses environmental context to measure the actual severity of a finding in four ways: as scoring inputs, through reachability, through language models and agents, and in industrial systems.

Scoring standards reserve a slot for environment-specific context and leave filling it to the consumer. CVSS modifies a base score with environmental metrics, such as the Modified Attack Vector, that an analyst supplies by hand for each deployment~\citep{first2019cvss31,first2023cvss40}. SSVC rates the exposure of the installed system~\citep{spring2021ssvc}, and the Exploit Prediction Scoring System (EPSS) excludes instance-specific context by design~\citep{first2026epssfaq}. Studies of environmental metrics simulate the gain of adding organizational context~\citep{fruhwirth2009context}, apply the metrics to published scores to study the resulting distribution~\citep{gallon2010environmental}, find them hard for assessors to apply on segmented networks~\citep{allodi2017environmental}, or propose deriving them from inventories and network topology~\citep{eitel2020environmental}. Context here is supplied by an analyst or an organizational model. The non-intrusive, context-based vulnerability scoring framework (NCVS) is the exception: it collects a cloud service's topology and configuration automatically and scores its vulnerabilities over the resulting dependency graph~\citep{zhuang2016ncvs}.

Reachability analysis reads context from the artifact or the network itself. For software dependencies, checking whether vulnerable code is deployed or called removes a large share of alerts. A manual study of 60 dependent projects found that as many as 73.3\% of those with outdated dependencies were not exposed~\citep{zapata2018smoother}, and about 20\% of vulnerable dependencies were never deployed~\citep{pashchenko2018counting}. Call-graph pruning cut npm audit false positives by 81\%~\citep{nielsen2021modular}, and fewer than 1\% of Maven packages have a call path to a vulnerable function in a dependency~\citep{mir2023transitivity}. Related analyses follow the same principle~\citep{ponta2018beyond,wu2023upstream}. For networks and cloud, attack graphs~\citep{sheyner2002automated,ou2005mulval} and automated reasoning over network and identity configuration~\citep{backes2019reachability,shevrin2023detecting} decide which weaknesses an attacker can reach. For rule-based cloud findings, controlled studies report that over 80\% of default public-access alerts on more than 1,000 storage buckets were false positives~\citep{dikshant2025precision}, and that active probes cut false positives by 93\% on average~\citep{dikshant2026alertnoise}. These works decide whether a finding applies; they do not re-grade how severe it is in its environment.

Language models first entered severity assessment without an environment. Predicting CVSS base metrics from descriptions is limited by descriptions that lack the deciding context~\citep{jafarikhah2025description}, and models asked for SSVC decisions from text over-predict risk~\citep{alhaddad2025prompting}. Environment-aware work supplies context in different forms. CVE-LLM, which evaluates Common Vulnerabilities and Exposures (CVE) entries with a large language model (LLM), reads a manufacturer's asset descriptions and vendor notifications and emits an affected or not-affected verdict with the CVSS environmental metrics. In production, with experts validating its output, it reaches a micro-F1 of 0.98 on those metrics~\citep{ghosh2025cvellm}. On VEX-Bench, a benchmark for Vulnerability Exploitability eXchange (VEX) assessments, agents read the downstream codebase to decide whether a supply-chain vulnerability affects it, and the strongest reach about 80\% F1 on that binary decision~\citep{shi2026vexbench}, while SastBench evaluates agentic triage of static-analysis findings~\citep{feiglin2026sastbench}. For cloud and Kubernetes, a judge-agent framework is validated on cloud-misconfiguration triage in large environments~\citep{garg2026jaf}, and KubeGuard reads runtime logs to propose least-privilege configurations~\citep{sgancohen2025kubeguard}. On a purpose-built cluster, giving a model the runtime topology lifts its correct patches, for Kubernetes findings that depend on that topology, to 78.0\% from 11.1\%~\citep{shaikh2026runtime}. Our earlier work adjusts finding severity with a model inside an asset-level scoring pipeline~\citep{engelberg2026ai}, finds that agents answer identity tasks more accurately when given structured relational context~\citep{yavin2026ispm}, and reports higher coverage than coding agents on cloud investigation tasks for a security context layer~\citep{goldberg2026coding}. None of these re-grades third-party cloud findings in production with evidence joined across products, or checks the deciding facts against the live environment.

Industrial platforms now ship agents for this task. Wiz's Green Agent traces posture issues through cloud resources, access policies, network reachability, and code changes, and returns a remediate-or-ignore verdict with suggested severity adjustments~\citep{wiz2026greenagent}; its Blue Agent issues verdicts on threat detections~\citep{wiz2026blueagent}. Upwind's Red Agent validates exploitability from runtime reachability and attack paths~\citep{upwind2026redagent}. CrowdStrike pairs a learned exploit-prediction model over threat intelligence with derived asset criticality~\citep{crowdstrike2025exprtai}, and Microsoft Defender for Cloud ranks recommendations by environment risk factors and attack paths~\citep{microsoft2026riskprio,microsoft2026attackpath}. Microsoft and Google also triage alerts with agents~\citep{msft2026alerttriage,google2026secopsagents}. Open-source projects follow the same pattern: a Prowler plugin lets a coding agent order posture findings by attack-path blast radius~\citep{prowler2026claudeplugin}, and GitHub Security Lab's taskflow agent triages code-scanning alerts with workflow and permission context~\citep{githubseclab2026taskflow}. None of these systems publishes a reproducible evaluation of whether its contextual re-rating is correct.

Together, these lines show that environmental context changes the actual severity of a finding, whether the context comes from analyst input, code and network reachability, or an agent's investigation. What is missing is an agent that gathers environment evidence for each finding across products, re-grades third-party severities in production, and is evaluated on whether the facts behind its verdicts hold.

\section{The Contextualization Agent}
\label{sec:agent}

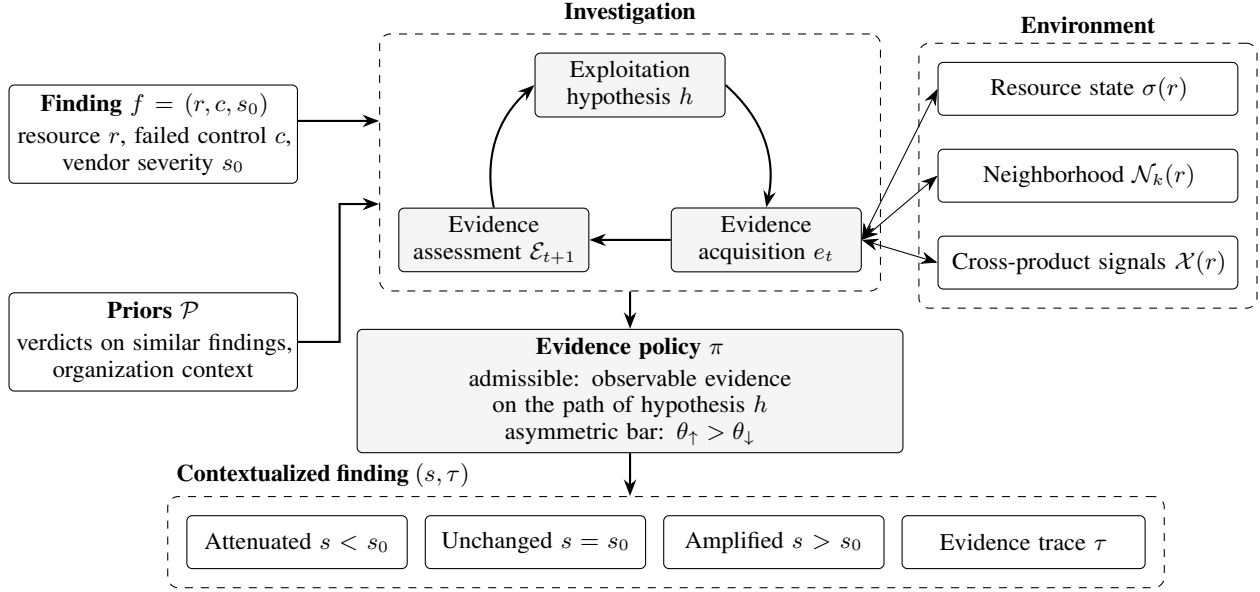
\begin{figure}[t]
	\centering
	\begin{tikzpicture}[
		font=\small,
		>={Stealth[length=2.2mm]},
		box/.style={draw, rounded corners=2pt, align=center, minimum height=0.7cm, inner sep=3pt, fill=white, execute at begin node={\hyphenpenalty=10000\exhyphenpenalty=10000}},
		stage/.style={box, text width=2.3cm, fill=gray!8},
		envbox/.style={box, text width=3.75cm},
		outbox/.style={box, text width=2.7cm},
		group/.style={draw, dashed, rounded corners=4pt, inner sep=7pt},
		gtitle/.style={font=\small\bfseries, anchor=south},
		flow/.style={->, thick},
		link/.style={<->, thin}
	]

	\node[box, text width=3.6cm, minimum height=1.3cm] (find) at (0.1, 1.75) {\textbf{Finding} $f=(r,c,s_0)$\\[2pt] \footnotesize resource $r$, failed control $c$,\\ \footnotesize vendor severity $s_0$};

	\node[stage] (hyp)  at (6.4, 2.4) {Exploitation\\ hypothesis $h$};
	\node[stage] (gat)  at (8.2, 0.35) {Evidence\\ acquisition $e_t$};
	\node[stage] (wei)  at (4.6, 0.35) {Evidence\\ assessment $\mathcal{E}_{t+1}$};
	\draw[flow] (hyp.east) to[bend left=30] (gat.north);
	\draw[flow] (gat.west) -- (wei.east);
	\draw[flow] (wei.north) to[bend left=30] (hyp.west);
	\node[group, fit=(hyp)(gat)(wei)] (gloop) {};
	\node[gtitle] at (gloop.north) {Investigation};

	\node[envbox] (e1) at (12.45, 2.35) {Resource state $\sigma(r)$};
	\node[envbox] (e2) at (12.45, 1.2) {Neighborhood $\mathcal{N}_k(r)$};
	\node[envbox] (e3) at (12.45, 0.05) {Cross-product signals $\mathcal{X}(r)$};
	\node[group, fit=(e1)(e3)] (genv) {};
	\node[gtitle] at (genv.north) {Environment};
	\draw[link] (gat.east) -- (e1.west);
	\draw[link] (gat.east) -- (e2.west);
	\draw[link] (gat.east) -- (e3.west);

	\coordinate (lfi) at ([yshift=0.55cm]gloop.west);
	\draw[flow] (find.east |- lfi) -- (lfi);

	\node[box, text width=3.6cm, minimum height=1.3cm] (prior) at (0.1, -1.0) {\textbf{Priors} $\mathcal{P}$\\[2pt] \footnotesize verdicts on similar findings,\\ \footnotesize organization context};
	\coordinate (lpr) at ([yshift=-0.55cm]gloop.west);
	\draw[flow] (prior.east) -- ++(0.55,0) |- (lpr);

	\node[box, text width=7.0cm, fill=gray!8] (pol) at (6.4, -1.65) {\textbf{Evidence policy} $\pi$\\[2pt] admissible: observable evidence on the path of hypothesis $h$\\ asymmetric bar: \mbox{$\theta_{\uparrow} > \theta_{\downarrow}$}};
	\draw[flow] (gloop.south) -- (pol.north);

	\node[outbox] (low) at (2.0, -3.65) {Attenuated $s < s_0$};
	\node[outbox] (unc) at (5.15, -3.65) {Unchanged $s = s_0$};
	\node[outbox] (rai) at (8.3, -3.65) {Amplified $s > s_0$};
	\node[outbox, text width=3.0cm] (trc) at (11.6, -3.65) {Evidence trace $\tau$};
	\node[group, fit=(low)(trc)] (gout) {};
	\node[gtitle] at (gout.north west) [anchor=south west] {Contextualized finding $(s,\tau)$};
	\draw[flow] (pol.south) -- (pol.south |- gout.north);
	\end{tikzpicture}
	\caption{Contextualization agent. $s_0$: vendor severity; $s$: contextualized severity; $\theta_{\uparrow}, \theta_{\downarrow}$: bars for raising and lowering a severity.}
	\label{fig:agent}
\end{figure}

The contextualization agent is a deep research agent in the sense of~\citep{huang2025deep}: it plans a multi-step investigation, retrieves over several sources, uses tools iteratively, and ends with a structured report, here a contextualized finding. Figure~\ref{fig:agent} shows its parts: the finding and its priors, the environment and the actions that reach it, the investigation, the evidence policy, and the contextualized finding.

A finding $f=(r,c,s_0)$ is a resource $r$ that fails a control $c$, with vendor severity $s_0$. The agent receives the definition of control $c$; for a third-party finding this is the vendor's name, description, and severity, without the check logic. It also receives the state and graph neighborhood of resource $r$, the other controls resource $r$ fails and passes, and findings from other tools that test for the same risk. The scope is technical: properties, configurations, relations, and guardrails. Our running example is a security group, resource $r$, flagged for admitting inbound traffic from any address on an administrative port, control $c$, with vendor severity $s_0$ = HIGH.

Evidence about resource $r$ comes in three kinds: the resource state $\sigma(r)$; the neighborhood $\mathcal{N}_k(r)$, holding the resources within $k$ hops of resource $r$ in the asset graph; and the cross-product signals $\mathcal{X}(r)$ held by other products. The evaluation attributes each verdict's decisive evidence to one of these kinds, and four actions reach them. A \emph{structured query} runs over the tabular store that unifies every connected product; it is the most complete source, reads the resource state $\sigma(r)$, and joins across products. A \emph{graph query} traverses several hops to return the neighborhood $\mathcal{N}_k(r)$, including the controls that neighboring resources fail. A \emph{catalog search} maps a domain-level request to the data other products hold and to how that data relates to resource $r$, reaching cross-product signals $\mathcal{X}(r)$ that the graph does not link. A \emph{neighborhood exploration} lists what a resource kind connects to, guiding the next graph query. For the security group, the resource state $\sigma(r)$ holds its ingress rules and attachments, the neighborhood $\mathcal{N}_k(r)$ the interfaces, load balancers, and instances behind it, and the cross-product signals $\mathcal{X}(r)$ what endpoint agents or vulnerability scanners report about those instances.

The investigation follows the reason-act pattern~\citep{yao2023react} over an exploitation hypothesis $h$ and an evidence set $\mathcal{E}_t$. At each iteration $t$ it makes three moves. It states or revises the exploitation hypothesis $h$: how the failure of control $c$ on resource $r$ would realistically be exploited. It acquires evidence items $e_t$ through the actions, in parallel when independent. It assesses them, adding them to the evidence set, $\mathcal{E}_{t+1}=\mathcal{E}_t\cup\{e_t\}$, recording what each item showed and an interim conclusion that directs the next step. The assessment also reads the priors $\mathcal{P}$: verdicts on similar findings and organization context. Verdicts on similar findings count as evidence, not as answers to copy, and a finding whose inputs have not changed reuses its earlier result. Organization context is of two kinds. A claim about the environment, such as a firewall fronting a service, is verified against the data before it applies; an accepted risk applies as stated. Only context that changed the conclusion is cited. The loop stops at a conclusion or when its budget runs out; in the second case the severity is unchanged. For the security group, the exploitation hypothesis $h$ is an attacker on the internet reaching a service behind it. A structured query shows that every attachment belongs to a load balancer, and a graph query shows that each load balancer is internal.

The evidence policy $\pi$ decides which evidence items count and how much evidence each direction needs. Its first rule is that an item counts only if it is observed in the data and bears on how the failure would be exploited, as stated by the exploitation hypothesis $h$. Encryption at rest, for example, does not offset a publicly readable bucket, because anyone allowed to read the bucket receives the decrypted data. A query that returns nothing is not evidence of safety, because the data may never have been collected, and a guardrail that could exist only in application code neither lowers nor raises the severity, because it cannot be observed. Its second rule is that raising a severity needs more evidence than lowering it, which we write as a higher bar for raising, $\theta_{\uparrow}$, than for lowering, $\theta_{\downarrow}$. Lowering requires one observed compensating property in the evidence set $\mathcal{E}_t$: for the security group, that every attached load balancer is internal. Raising requires positive evidence that the failure is worse here, such as an attached instance with a public address running a service that a scanner reports as critically vulnerable, together with an explicit check that no firewall, allowlist, or other plausible guardrail sits in front of it. When neither bar is met, the severity stays unchanged.

The output is the contextualized finding $(s,\tau)$: a contextualized severity $s$ and an evidence trace $\tau$. The contextualized severity $s$ is \emph{attenuated}, below the vendor severity $s_0$, in one of two grades, \emph{mitigated} when a property offsets the failure entirely (to LOW or informational) and \emph{reduced} when it compensates partially (one level down). It is \emph{amplified}, one level above the vendor severity $s_0$, when a property materially worsens this specific failure, and \emph{unchanged} otherwise. The evidence trace $\tau$ records the acquisitions and assessments and marks the key evidence; in the evaluation, LLM judges label this trace with a recalibration pattern. It also carries a technical rationale, a plain-language explanation, and a confidence that rates the strength of the evidence set $\mathcal{E}_t$ regardless of direction and stays low when the investigation is shallow or the deciding evidence would be unobservable. The security group remains reachable from inside the network, so the compensation is partial and the finding is reduced from HIGH to MEDIUM, with the attachment and load-balancer queries marked as key evidence.

\section{Evaluation}
\label{sec:evaluation}

\paragraph{Criteria.} We measure three criteria on a production corpus. \emph{Faithfulness} asks whether the fact a verdict rests on is true in the live environment, in the spirit of atomic factual precision~\citep{min2023factscore}. Of the live findings whose decisive fact the probes can decide under the scope rules of the live check, the set $D$, let $C$ be those whose fact matches the live state. Faithfulness is the share that matches, $\Phi = |C|/|D|$. Because some findings cannot be checked, we also report two variants: the inclusive faithfulness $\Phi_R$, computed over every finding the probes could read, and the lower-bound faithfulness $\Phi_L$, computed over all live findings with every unchecked finding counted as wrong.

\emph{Context dependence} asks how often the deciding evidence comes from beyond the flagged resource. For each finding $f$ in the traced population $T$, let $\ell(f)$ record which kind of evidence decided it, as given by the retrieval label of the evidence attribution: the resource state $\sigma(r)$ of the flagged resource $r$, its $k$-hop neighborhood $\mathcal{N}_k(r)$, or the cross-product signals $\mathcal{X}(r)$ defined in Section~\ref{sec:agent}. Context dependence is the share of findings decided by anything other than the resource state, $\Delta = |\{f \in T : \ell(f) \neq \sigma\}|/|T|$. We report it per kind of evidence, and add a necessity share for rules that move in both directions.

\emph{Regularity of reasoning} asks whether the verdicts follow a small set of recurring patterns that independent LLM judges, separate runs of a language model applying the same codebook~\citep{tai2024examination}, identify the same way. We measure the first part by saturation, the point after which further verdicts add no new pattern~\citep{guest2006how}, and the second by Krippendorff's $\alpha$~\citep{krippendorff2004content}, an agreement coefficient that is 1 for perfect agreement and 0 for chance, between the judges' runs. Neither measures whether the agent would repeat its verdicts, since the agent was not re-run.

\paragraph{Corpus.} The corpus is the HIGH queue of both platforms in eight real production environments: 9,967 findings, 4,222 raised by Platform A and 5,745 by Platform B. Each finding is credited, by rule namespace, to the platform that raised it, and not to any tool merely connected to the same environment. The unit of analysis is the finding, one (environment, rule, resource group) record in which near-identical resources in the same situation share one verdict. All Platform A findings come from four of the eight environments, and 92\% of the findings the agent re-graded are on Amazon Web Services (AWS), with the rest on a software-as-a-service identity provider (580) and Google Cloud (9). Results refer to three subsets of the re-graded findings: the \emph{traced population}, the 7,243 whose investigation trace was recovered and which the LLM judges label; the \emph{characterized corpus}, the 4,772 traced findings whose decisive fact a live cloud application programming interface (API) can check; and the \emph{residual}, the remaining 2,471, whose decisive facts sit in other products and policies.

\paragraph{Procedures.} Four procedures turn the agent's verdicts into the three criteria. \emph{LLM-judge labeling} gives every verdict three labels: a \emph{pattern}, the kind of reasoning step that decided it, such as ``not internet-reachable''; a \emph{location}, whether its decisive fact sits on the flagged resource or elsewhere; and a \emph{retrieval}, whether that fact came from the resource itself, a graph traversal, or another product's data. The judges are independent runs of Claude Opus 4.8 that read each investigation trace with a fixed codebook, which defines the labels, and a rubric, which settles ambiguous cases in favor of on-resource; the patterns themselves were first derived from the traces by grounded theory~\citep{strauss1990basics,glaser1967discovery}. A second, independent judge run labels a sample of the same verdicts again~\citep{tai2024examination}, and the agreement between the two judge runs, Krippendorff's $\alpha$ with Cohen's $\kappa$~\citep{cohen1960coefficient} and raw agreement alongside, measures how reproducibly the agent's reasoning falls into the patterns, the second part of regularity of reasoning. \emph{Evidence attribution} uses the location and retrieval labels to compute context dependence $\Delta$. Because the agent was not re-run, the \emph{necessity analysis} tests from the existing verdicts, as an observational study would~\citep{rosenbaum2002observational}, whether that outside evidence actually set a verdict's direction: a rule that the agent lowers on some findings and raises on others checks the same condition on all of them, so where the decisive fact is off-resource, that outside fact set the direction. \emph{Live verification} gives faithfulness $\Phi$. Read-only probes on five resource types re-read each decisive fact, for example whether the running example's load balancers are still internal, directly from the cloud API instead of the asset graph the agent used, so an error in the graph cannot confirm itself; a verdict is confirmed if the fact matches it and contradicted otherwise~\citep{apa1954technical,cronbach1955construct}. The probes run on eight AWS accounts where we hold read-only access, and since only Platform B monitors these accounts, faithfulness $\Phi$ is measured on Platform B findings. Under the scope rules, each resource is checked in one direction, as an amplification if any finding on it was amplified and otherwise as an attenuation, and every rate carries a Wilson score interval~\citep{wilson1927probable,brown2001interval}.

\hyperref[box:example]{Box~1} traces one finding through the investigation and the three procedures.

\begin{tcolorbox}[enhanced, float=tb, floatplacement=tb, colback=white, colframe=black!75, colbacktitle=black!80, coltitle=white, boxrule=0.6pt, arc=1.5pt, left=6pt, right=6pt, top=4pt, bottom=4pt, toptitle=1.5pt, bottomtitle=1.5pt, fonttitle=\bfseries\small, title={Box 1.\ One finding, from vendor grade to live verification (illustrative)}]
\phantomsection\label{box:example}\footnotesize
\newcommand{\boxhead}[2]{\par\noindent{\color{black!80}\rule[-1pt]{2pt}{8pt}}\hspace{4pt}\textbf{\textsc{#1}}\hfill{\color{black!60}\textit{#2}}\par\vspace{2pt}}
\newcommand{\sev}[2]{\colorbox{#1}{\rule[-1.5pt]{0pt}{8pt}\textbf{#2}}}
\boxhead{1\quad Finding $f=(r,c,s_0)$}{as raised by Platform B}
\begin{tabularx}{\linewidth}{@{}l X@{}}
resource $r$ & container service \texttt{svc-payments-worker} \\
control $c$ & runtime exposure: the workload's image carries a package with a known exploitable vulnerability \\
vendor severity $s_0$ & \sev{red!18}{HIGH}\ \ the grade this rule carries in every environment \\
\end{tabularx}

\vspace{3pt}
\boxhead{2\quad Investigation}{exploitation hypothesis $h$: an attacker exploits the vulnerable package in a running task}
\begin{tabularx}{\linewidth}{@{}l l >{\raggedright\arraybackslash}X >{\raggedright\arraybackslash}X@{}}
\toprule
step & evidence source & action & observation \\
\midrule
$e_1$ & resource state $\sigma(r)$ & structured query: desired and running task count of $r$ & \texttt{desired=0}, \texttt{running=0}: scaled to zero \\
$e_2$ & neighborhood $\mathcal{N}_k(r)$ & graph traversal: what routes traffic to $r$ or starts its tasks & no load-balancer target, no scheduled start \\
\bottomrule
\end{tabularx}\\[2pt]
\textit{Evidence policy $\pi$.} With no running task, nothing on the path of $h$ executes the vulnerable package: the exposure is not realized. The service can be scaled up again, so the compensation is partial and the grade drops one level. In another environment the same rule is raised to CRITICAL, where the workload runs and the vulnerable package is actively used.

\vspace{3pt}
\boxhead{3\quad Contextualized finding $(s,\tau)$}{evidence trace $\tau$ marks $e_1$ as key evidence}
\sev{red!18}{HIGH}\ $\longrightarrow$\ \sev{orange!25}{MEDIUM}\quad reduced.\quad Decisive fact: \emph{the service runs no task}.

\vspace{4pt}
\boxhead{4\quad Evaluation}{how this verdict enters each criterion}
\begin{tabularx}{\linewidth}{@{}l >{\raggedright\arraybackslash}X l@{}}
\toprule
procedure & result on this finding & criterion \\
\midrule
live verification & read-only cloud API read of the service's running task count, bypassing the asset graph: still \texttt{0}\ \ $\Rightarrow$\ \ \textcolor{green!40!black}{\textbf{confirmed}}; a running task would count as a contradiction & faithfulness $\Phi$ \\
evidence attribution & decided by the resource state $\sigma(r)$, so it does not count toward $\Delta$; the fact was on the resource, yet the vendor grade ignored it & context dependence $\Delta$ \\
necessity analysis & the rule moves both ways, but the decisive fact is on the resource, so the finding is not in the necessity share & necessity share \\
LLM-judge labeling & pattern \emph{no live service}, location \emph{on-resource}, retrieval \emph{structured query}; labeled again by a second judge run & regularity, $\alpha$ \\
\bottomrule
\end{tabularx}
\end{tcolorbox}

\section{Results}
\label{sec:results}

Of the 9,967 HIGH findings, the agent re-grades three in four (75.1\%): 58.5\% down and 16.6\% up, while 2,478 (24.9\%) keep the vendor's grade. Among the re-graded findings of the characterized corpus (Table~\ref{tab:transitions}), a single vendor grade spreads across the whole scale, from informational to CRITICAL.

\begin{table}[t]
	\caption{Contextualized grade of vendor HIGH findings ($n = 4{,}772$).}
	\label{tab:transitions}
	\centering
	\small
	\begin{tabular}{lrrrrl}
		\toprule
		Grade    & Count & Share  & A  & B   & Move \\
		\midrule
		MEDIUM   & 3,148 & 66.0\% & 47.8\% & 74.6\% & down \\
		CRITICAL & 950   & 19.9\% & 26.9\% & 16.6\% & up \\
		LOW      & 555   & 11.6\% & 19.1\% & 8.1\%  & down \\
		INFO     & 119   & 2.5\%  & 6.2\%  & 0.7\%  & down \\
		\bottomrule
	\end{tabular}
\end{table}

Six recurring patterns account for the characterized verdicts. Five mainly lower severity: the resource is inert, attached to nothing or orphaned (32.3\%); it is not internet-reachable (25.4\%); no live service runs behind it (12.4\%); it admits only specific sources (7.5\%); or an account-level block neutralizes the condition (3.0\%). One mainly raises it: a public exposure that is realized (19.4\%). Each pattern moves severity in the direction that established frameworks prescribe for its fact, such as the CVSS attack vector~\citep{first2019cvss31}, SSVC's system exposure~\citep{spring2021ssvc}, attack-graph reachability~\citep{ou2005mulval}, and likelihood conditioned on deployed controls~\citep{nist2012sp80030}. Verdicts in the residual turn mainly on compensating controls and private endpoints recorded in other products and policies. The patterns attach to findings, not to rules: 62 of the 138 rules lower some findings and raise others, 57 of them inside a single environment and none across resource types. A security-group rule, for example, is lowered when the group fronts only internal load balancers and raised when the environment shows an internet-facing path.

\begin{table}[!htb]
	\caption{Evaluation results. Brackets: 95\% Wilson interval; for necessity, bounds from label disagreement.}
	\label{tab:criteria}
	\centering
	\small
	\begin{tabular}{l l r}
		\toprule
		Criterion & Measure & Value \\
		\midrule
		\multirow{3}{*}{Faithfulness} & Confirmed live, $\Phi$ & 99.4\% [99.1, 99.6] \\
		 & \quad inclusive, $\Phi_R$ & 97.3\% [96.7, 97.7] \\
		 & \quad lower bound, $\Phi_L$ & 81.8\% [80.7, 82.8] \\
		\midrule
		\multirow{4}{*}{Context dependence} & Beyond the flagged resource, $\Delta$ & 49.0\% \\
		 & \quad graph neighborhood & 29.8\% \\
		 & \quad other products & 19.2\% \\
		 & Necessity share & 29.6\% [27.3, 31.9] \\
		\midrule
		\multirow{3}{*}{Regularity of reasoning} & Judge agreement, pattern, $\alpha$ & 0.90 \\
		 & Judge agreement, location, $\alpha$ & 0.94 \\
		 & Saturation & 22\% of traces \\
		\bottomrule
	\end{tabular}
\end{table}

Table~\ref{tab:criteria} reports the three criteria, starting with faithfulness, the primary measure.
\begin{itemize}
	\item \emph{Faithfulness.} The live cloud state confirms the decisive fact of 99.4\% of the verdicts the probes can decide (faithfulness $\Phi$), including all 181 decided exposure amplifications, and of 528 of 531 resources when each resource counts once. The 24 contradictions sit on three resources whose live state differs from the verdict. Inclusive faithfulness $\Phi_R$, over every finding the probes could read, is 97.3\%, and lower-bound faithfulness $\Phi_L$, which counts every unchecked finding as wrong, is 81.8\%. The check covers Platform B findings, tests the decisive fact rather than the grade, and does not reach the residual.
	\item \emph{Context dependence.} About half of the verdicts rest on evidence beyond the flagged resource (context dependence $\Delta = 49.0\%$), drawn from its graph neighborhood (29.8\%) or from another product's signals (19.2\%). A rater that reads only the flagged resource therefore has the decisive evidence for at most 51.0\% of the verdicts, and one that reads a single product's graph for at most 80.8\%. For 29.6\% of the traced population (the necessity share), the finding falls under a rule that moves both ways and an off-resource fact decided which way. Outside evidence concentrates in attenuations, where showing a finding safe means ruling out every path to it.
	\item \emph{Regularity of reasoning.} Two independent judge runs agree on the pattern (Krippendorff's $\alpha = 0.90$, Cohen's $\kappa = 0.90$, 91.2\% raw agreement) and, under the on-resource rubric, on the location label ($\alpha = 0.94$), and no new pattern appears after the first 22\% of traces. The verdicts thus reduce to a small set of patterns that are identified reproducibly.
\end{itemize}

\section{Discussion}
\label{sec:discussion}

The results support treating severity as a property of a finding in its environment rather than of the rule that raised it. Because severity orders remediation, re-grading three in four of the vendor's HIGH findings changes their place in a severity-ordered queue. The same rule moving both ways inside one environment, on resources of one type, places the deciding variation at the level of the individual resource, below the level at which a rule rewrite or an environment-wide policy would operate. The three criteria qualify this reading. High faithfulness indicates that the re-grades rest on facts that hold in the live environment, for the exposure facts the probes can read. Context dependence locates those facts: about half lie beyond the flagged resource, in two tiers, graph evidence that a product could hold but its per-rule severity does not use, and evidence that exists only when products are joined. Regularity of reasoning indicates that the re-grades follow a small set of reproducible patterns whose directions match established frameworks, so each verdict can be explained and checked.

We argue that verified, environment-specific reading of each finding is a precondition for acting on findings autonomously. Reading and acting differ in blast radius, reversibility, and authorization: without an analyst in the loop, acting on a false HIGH disrupts production, and failing to act on a true CRITICAL leaves an exposure open. This study measures reading only: whether each verdict's decisive fact holds live, where that fact lies, and whether the reasoning falls into reproducible patterns, with the facts read through the same interface an automated responder would use before acting. It does not measure any action.

Several limitations bound the scope. The study covers one system's verdicts on the HIGH queue of two platforms in eight production environments, mostly on AWS, so the patterns transfer as a set, not as fixed frequencies, and findings the environment would raise from MEDIUM or LOW are not observed. Faithfulness is measured on one platform and five resource types, against the decisive fact rather than the grade, and recall, the deciding facts the agent missed, is not measured. The labels come from runs of a single model, and without incident labels the study cannot show that findings graded up are more likely to be exploited, or findings graded down less likely, than their vendor grade implies.

\section{Conclusion}
\label{sec:conclusion}

A contextualization agent, a deep research agent over a cross-signal asset graph, re-grades three in four of 9,967 third-party HIGH findings, mostly downward, and the same rule often moves in both directions within one environment. Its verdicts are faithful to the live environment, with 99.4\% of the decisive facts the probes can decide confirmed (81.8\% as a lower bound); they depend on evidence beyond the flagged resource for about half of the findings, a fifth only in another product's signal; and they follow six reproducible patterns shared by two platforms. The rule-severity gap is large in production, and closing it requires evidence from the whole environment.

\bibliographystyle{unsrtnat}
\bibliography{references}

\end{document}